# Gene-Environment Interaction: A Variable Selection Perspective

Fei Zhou[1], Jie Ren[1], Xi Lu[1], Shuangge Ma[2] and Cen Wu[1]

1. Department of Statistics, Kansas State University. Manhattan, KS 66506, USA
2. Department of Biostatistics, School of Public Health, Yale University, New Haven, CT 06510, USA

Correspondence: wucen@ksu.edu

Abstract:

Gene-environment interactions have important implications to elucidate the genetic basis of complex diseases beyond the joint function of multiple genetic factors and their interactions (or epistasis). In the past, G × E interactions have been mainly conducted within the framework of genetic association studies. The high dimensionality of G × E interactions, due to the complicated form of environmental effects and presence of a large number of genetic factors including gene expressions and SNPs, has motivated the recent development of penalized variable selection methods for dissecting G × E interactions, which has been ignored in majority of published reviews on genetic interaction studies. In this article, we first survey existing overviews on both gene-environment and gene-gene interactions. Then, after a brief introduction on the variable selection methods, we review penalization and relevant variable selection methods in marginal and joint paradigms respectively under a variety of conceptual models. Discussions on strengths and limitations, as well as computational aspects of the variable selection methods tailored for G × E studies have also been provided.

1. Introduction

Gene× Environment (G × E) interactions, in addition to the genetic and environmental main effects, have important implications for elucidating the etiology of complex diseases, such as cancer, type 2 diabetes and cardiovascular diseases ([1−5]). Multiple G × E studies have shown that the genetic contribution to the variation in disease phenotype or increase in disease risks are often mediated by environmental effects. Historically, G × E interactions have been examined from the perspective of assessing specific genetic effect under dichotomous environmental exposures ([6]). With the availability of high-density genetic polymorphisms such as single nucleotide polymorphisms (SNPs), it has become possible to establish the statistical associations between millions of genetic variants and disease status or phenotype in genetic association studies ([7,8,9]), which has also made investigation of G × E interactions possible at the more comprehensive human genome scales ([10,11,12]).

The dissection of G × E interactions in genetic association studies, such as genome wide association study (GWAS), has been mainly conducted based on the assessment of statistical significance. For example, in the genome wide case-control association studies of type 2 diabetes, with body mass index (BMI) as the environmental factor, the significance of the interaction between BMI and each one of the SNPs can be evaluated using p-values from the marginal test accounting for the interaction ([10]). After multiple test adjustment, important interaction effects can be identified when the signals are beyond the genome-wide significance level.

Furthermore, the genetic association studies can be understood from a related but distinct perspective. Consider the data matrix where the columns are corresponding to features (or variables), such as all the main and interaction effects in a G × E study, and rows are corresponding to samples (or observations). As the number of columns is usually much larger than the sample size in a typical G × E interaction study, the data matrix is of "large dimensionality, small sample size" nature. Thus, the central statistical task is to hunt down the subset of important main and interaction effects that is associated with the disease outcome, which can be reformulated as a high dimensional variable selection problem in the regression framework. Specifically, the regression coefficients of variables (representing main and interaction effects) are continuously shrunk towards zero. A zero coefficient after shrinkage denotes that the corresponding effect is not included in the final model, and has no association with the response, such as the disease phenotype. Therefore, variable selection can be performed with parameter estimation simultaneously. Such a variable selection method is referred as penalization or regularization ([13,14]).

Generically, penalized regression coefficients can be defined as

$$\hat{\beta} = \text{argmin}_\beta \{L(D;\beta) + P(\lambda;\beta)\},$$

where $L(D;\beta)$ is a loss function based on the observed data $D$ and regression coefficients $\beta$ to quantify the lack-of-fit. It can be a least square loss function or a negative log-likelihood function. The penalty function, $P(\lambda;\beta)$, measures the model complexity with tuning parameter $\lambda$. As $\lambda \to +\infty$, larger amount of penalty is imposed on $\hat{\beta}$, and more components of $\hat{\beta}$ become zeros. Accordingly, fewer features will be included in the final model. The phenomena of zeros in $\hat{\beta}$ is termed as sparsity in the literature of penalized variable selection. On the other hand, when $\lambda \to 0$, the model becomes more complex since more features are included in the final model, Tuning parameter $\lambda$ balances the tendency towards two extremes. A properly tuned $\lambda$ will lead to a reasonable number of variables with satisfactory interpretability and superior prediction performance.

Recently, penalization methods have been extensively developed for G × E studies. However, our limited literature search indicates that very few of the published review articles in genetic interaction analysis, including both G × E and G × G interactions, has included the penalization methods in the toolbox for interaction studies. Furthermore, none of the surveys conducted so far has investigated G × E studies from the viewpoint of variable selection. In Table 1, we provide a partial list of existing reviews on genetic interaction studies ([1,2,4,5,6,10,12,15−33]). We have found that none of the methodological based review papers is focused on G × E interactions. For example, McKinney et al. ([24]), Koo et al. ([30]), Wei et al. ([32]) and Niel et al. ([33]) survey machine learning and other methods for detecting gene-gene interactions. Among them, Niel et al. ([33]) has briefly brought up the concept of high dimensional data in the context of epistasis studies, and pointed out that penalized regression methods can be adopted to detect SNP-SNP interactions. However, they have concluded that penalization techniques are subject to severe limitations and have not pursued reviewing interaction studies from the perspective of variable selection.

The recent development of a large amount of powerful penalization methods have overcome the obstacles and made their applications in interaction studies, especially the G × E studies, successful. It is therefore urgent to conduct a methodological survey of G × E interaction studies by prioritizing the role of (penalized) variable selection since such a new trend has not been systematically reflected in existing survey articles on this topic so far. To keep this review self-contained, we will also include methods that are closely related to penalization in interaction analyses, such as the significance based and Bayesian studies, and investigate the G × E studies within the unified framework of variable selection. Such an arrangement will provide a unique viewpoint of examining the interaction studies. We conduct a deep review in penalization and relevant variable selection methods with a narrow but potentially important scope. Due to the limitation of our survey, a wide category of existing important methods for genetic interaction studies have not been included. For topics frequently discussed in general surveys of G × E and epistasis studies, including study designs, power and sample size of the study, interpretation of biological and statistical interactions, we refer readers to the overviews summarized in Table 1.

Table 1. Reviews on Genetic Interaction Studies (a partial list).

| Reference | Type | Description |
|---|---|---|
| **Hunter** ([1]) | G × E | Reviews the descriptions (based on qualitative models, statistical models and biological plausibility), study designs (family based studies, association study in unrelated individuals, retrospective, prospective and case-only association studies) and technical challenges (sample size) and some applications of G × E interactions. |
| **Simonds et al.** ([2]) | G × E | Systematically surveys published articles in gene-environment interaction studies from two relevant databases from January 1, 2001, to January 31, 2011. Results include the most frequently examined complex diseases and environmental factors in G × E studies. |
| **Cornelis et al.** ([4]) | G × E | Summarizes recent approaches, studies (study design and selecting gene and environment) and continued challenges (balancing type 1 error and statistical power, measuring the environment and selecting genes), emerging approaches and future perspectives of G × E interactions. |
| **Dempfle et al.** ([5]) | G × E | The main points include (1) potential applications of G × E interactions in the area of clinical care and public health, (2) definition and meaning of interaction, (3) study design, power and sample size, and (3) methodological challenges and perspectives |
| **Ottman** ([6]) | G × E | Discuss the definition of G × E interactions, how to model the relationship between genotype and exposure and how to test the models. |

| **Cornelis et al. ([10])** | G × E | Systematically conducts 7 statistical tests (standard case-control, case-only, semiparametric MLE, empirical Bayesian shrinkage, two-stage, joint 2df, semi-MLE 2 df) on two case control GWAS, the Health Professionals Follow-up Study (HPFS) and the Nurses' Health Study (NHS) |
|---|---|---|
| **Winham and Biernacka ([12])** | G × E | Besides reviewing study design and conventional analytical methods for G × E interactions, the article also discusses new directions including data mining methods for interaction studies and gene- and pathway-level G × E analysis. |
| **Caspi and Moffitt ([15])** | G × E | Provides a unified framework to integrate neuroscience and gene-environment interaction research along the line of neuroscience evidence base, epidemiological G × E research, and experimental neuroscience. Also discusses nature and nurture in G × E studies. |
| **Thomas ([16])** | G × E | This is a comprehensive review on G × E interaction studies including challenges (exposure assessment, power and sample size, heterogeneity and replication), study designs (basic epidemiologic designs, hybrid designs, family based designs and GWAS designs) and strategies of mining GWAS data for G × E studies, as well as experimental validation. |
| **Ober and Vercelli ([17])** | G × E | Reviews gene-environment interactions by especially focusing on asthma as a model disease. Environmental exposures unique to asthma, including environmental tobacco smoke (ETS) exposure and maternal asthma, have been discussed. |
| **Fletcher and Conley ([18])** | G × E | This article provides discussions and examples for G × E interaction studies from the perspective of social science (causal inference in particular). Examples include stress by genotype interaction under depressive phenotypes, and genotype by risky behavior interaction under health phenotypes in social sciences. |
| **McAllister et al. ([19])** | G × E | Highlights issues and main themes in gene-environment interaction studies, including a brief survey of analytical methods, environmental exposure assessment, functional validation and discovery and examples from human population studies under both Mendelian-like traits and complex diseases. |
| **Wu and Ma ([20])** | Both | Review robust genetic interaction analysis methods for both marginal and joint analysis, to address (1) model mis-specification, (2) outliers/contaminations in the response, and (3) outliers/ contaminations in predictors. |
| **Cordell ([21])** | G × G | Provides a historical background to the study of epistatic interaction effects and discusses a number of commonly used definitions and interpretations of epistasis. The mathematical formulation and statistical methods to detect epistasis have also been discussed. |
| **Moore ([22])** | G × G | Formulates a working hypothesis that epistasis is an essential building block of the genetic basis of complex disease and that complex interactions are more critical than independent main genetic effects. Also introduces multifactor dimensionality reduction to detect gene-gene interactions. |
| **Moore ([23])** | G × G | A conceptual paper on epistasis. Genetical, biological and statistical epistasis have been discussed. |
| **McKinney et al. ([24])** | G × G | Reviews popular machine learning methods, such as neural networks, cellular automata, random forests, and multifactor dimensionality reduction, to detect epistasis. A flexible and comprehensive framework for data mining and knowledge discovery through integrating MDR with other machine learning methods has been discussed. |
| **Phillips ([25])** | G × G | The importance of epistasis has been examined by considering high-throughput functional genomics, system biology approaches and pursuing the genetic architecture of evolution at the levels of specific molecular changes. Different perspectives on gene-gene interaction have been discussed, including functional, compositional and statistical epistasis. The role of gene-gene interaction in dissecting regulatory pathways and genetic mapping of complex diseases has been discussed. |
| **Cordell ([26])** | G × G | Surveys the methods and associated software packages available for the detection of epistasis in human genetic diseases. Key concepts related to statistical interaction have been discussed. This review examines regression based tests of interaction, exhaustive search and data mining methods in the context of gene-gene interactions. In particular, Bayesian model selection techniques are pointed out as an effective approach for epistasis studies. |

| Moore and Williams. ([27]) | G × G | In addition to investigating critical concepts related to epistasis, this review argues it as a ubiquitous component of the genetic architecture of complex diseases. Challenges include modelling interaction nonlinearly and interpretation of the identified effects. Thoughts on implications of G×G studies for personal genetics and recommendations for the improvement have been provided. |
|---|---|---|
| Wang et al. ([28]) | G × G | This study investigates the fundamental meaning of interaction through both the statistical and biological perspectives. Alternative meanings of interaction, including additive, multiplicative, quantitative and synergistic interactions, have also been provided. The relationship of interaction to the magnitude of measurement, unbalanced two-way table and gametic phase disequilibrium has also been addressed. |
| Li et al. ([29]) | G × G | This study systematically surveys statistical methods for detecting gene-gene interactions under a variety of phenotypic traits, including quantitative and survival traits. In particular, methods for unrelated case-control study and family based case control study have been surveyed. |
| Koo et al. ([30]) | Mostly on G × G, also including G × E | This article first introduces three types of synthesis, epistasis and suppression. The development and applications of major machine learning methods, neural networks, support vector machine and random forests, as well as their variants, in gene-gene interaction studies have been carefully explored. The strength and limits of these methods have been investigated. |
| VanderWeele et al. ([31]) | G × G and G × E | This tutorial provides a comprehensive introduction to the interaction between effects of exposures. Besides concepts and motivations for interaction studies, interactions on both additive and multiplicative have been examined under different statistical models. Study designs and properties of interaction analysis, as well as limitations and extensions have also been discussed. |
| Wei et al. ([32]) | G × G | This is a methodological review for epistasis studies. Methods for detecting gene-gene interactions investigated in the study include regression based, LD based, Bayesian, data filtering and machine learning methods. Discussions on the relevance of epistasis in GWAS and potential pitfalls in interpreting statistical interactions have also been provided. |
| Niel et al. ([33]) | G × G | A broad spectrum of methods for epistasis detection have been surveyed according to different strategies, including exhaustive search, two-stage approaches and non-exhaustive searches enhanced by machine learning. |

2. Method

Marginal and joint analyses are the two paradigms for Gene × Environment interaction studies ([20,26]). In marginal analysis, the interaction between one or a small number of omics features (such as gene expressions or SNPs) and E factors are considered at one time. In joint analysis, E factors and a large number of omics features are analyzed in a single model. As only a subset of the main and interaction effects is expected to be associated with the phenotype, variable selection plays an important role in both paradigms.

2.1 The two paradigms: marginal and joint analysis

**Marginal analysis based variable selection**: Variable selection through marginal analysis is usually conducted based on statistical significance. Consider a commonly adopted conceptual model for G × E interaction studies:

$$\text{Outcome} \sim \text{Cs} + E + G + G \times E, \qquad (1)$$

where the outcome variable can be continuous disease phenotypes, categorical disease status, or (censored) patients' survival. With a slight abuse of notation, let E and G represent one environmental variable and one genetic variable, respectively, and denote Cs as multiple clinical variables. In a GWAS, a marginal regression model can be fitted with respect to one G factor (SNP) at a time across the whole genome. Selection of significant interactions can be accomplished based on marginal p-values (readily calculated from existing statistical software) or likelihood. The most prominent advantage of marginal analysis is its computational convenience and conceptual simplicity. It only requires software implemented with standard procedures in general. Therefore, it is still popular nowadays especially for large scale studies ([10,11,12]).

A common objective of both the genetic association studies and penalized variable selection methods discussed in this article is to "find a needle in a haystack", or search for the signals associated with the clinical outcome of interest from a large amount of noisy ones. Therefore, we include both in the variable selection framework although the two are distinct methodologically.

**Joint analysis based variable selection**: Unlike the marginal analysis which conducts variable selection and model building in successive steps, the joint analysis achieves the two simultaneously. Within the joint analysis framework, variable selection has been mainly developed based on the following techniques:

1. Penalization. As discussed in Introduction, this family of approaches seeks the subset of important features through continuously shrinking the regression coefficients to 0, with the nonzero coefficients corresponding to selected features. Representative baseline penalization methods include LASSO (least absolute shrinkage and selection operator), SCAD (smoothly clipped absolute deviation), adaptive LASSO, and MCP (minimax concave penalty) ([34−38]), where shrinkage has been imposed on individual coefficients without considering the interconnections among the features. Penalization approaches beyond the baseline level have been developed to accommodate complex structures among features , including elastic net ([38]) and network based penalty ([39,40]) for correlation, group LASSO for group structure rising from gene set or pathways ([41,42]), sparse group LASSO or sparse group MCP for bi-level selection ([43]). In this article, we will provide detailed and in-depth discussion on how tailored penalization methods have been developed for a variety of G × E interaction studies.

2. Bayesian variable selection. Bayesian variable selection can be classified into the following four groups: indicator model selection, stochastic search variable selection, adaptive shrinkage and model space approach ([44]). Among them, adaptive shrinkage has deep connections to penalization. For instance, LASSO and group LASSO can be formulated within the Bayesian framework by assigning univariate and multivariate independent and identical double exponential priors to regression coefficients on individual and group level, respectively ([45,46]). Existing Bayesian variable selection methods for G × E interaction studies, including [47−50], have been mainly proposed under adaptive shrinkage and indicator model selection ([51,52,53]).

3. Other variable selection methods. There is a diversity of variable selection methods that are potentially applicable for G × E studies. For example, Boosting, a popular machine learning method, aggregates multiple weak learners (individual features of weak predictive power for the response variable) into a strong learner (a model of strong predictive power) ([54,55,56]). Within a regression framework, boosting has strong connections to penalization ([57−59]), which makes it a natural choice for detecting important G × E interactions ([60,61]). Support vector machine, another popular machine learning technique which is tightly connected to penalization in the form of "hinge loss + ridge penalty", can also be adopted for G × E interactions ([62,63]). Despite success in these studies, majority of the machine learning methods have been developed for epistasis studies ([24,30,32, 33]).

Our discussion of marginal and joint analysis based variable selection methods does not necessarily imply that they only belong to joint (or marginal) paradigm and cannot accommodate marginal (or joint) analysis. For example, penalization methods have already been developed for marginal analysis (see a detailed discussion of Section 2.4.2), and significance based variable selection has also been proposed in joint G × E studies (Section 2.4.1). Besides, as penalization generally works for moderately high dimensional data, it is a common practice to first carry out marginal screening to reduce the number of features subsequently analyzed using penalization. The validity of coupling marginal screening with joint penalized variable selection depends on the highly challenging theory of sure independence screening ([64−66]). Such a theory demands that omics features relevant to the phenotype are only in weak correlation with those "noisy" features, which is not likely to be true from a biological perspective.

Unsupervised approaches, including principal component analysis (PCA), clustering, canonical correlation analysis (CCA), partial least square (PLS), among many others, are closely related to variable selection. The response variable is not (fully) available in unsupervised analysis. Take PCA as a representative example. It can be interpreted as a ridge regression with PC loadings (which are usually nonzero) being denoted by the regression coefficients, which has the form of "least square loss + ridge penalty" ([67]). To improve model fitting and interpretability, sparse PCA shrinks the nonzero PC loadings toward zero, with a formulation of "least square loss + ridge penalty + L1 penalty", or equivalently, "least square loss + elastic net penalty" ([67]). Therefore, both PCA and its sparse counterpart share the spirit of "unpenalized loss function + penalty function" formulation from penalization. Please refer to Wu et al. ([68]) for a detailed discussion of unsupervised analysis and variable selection under this formulation. PCA usually yields results that are difficult to interpret. For instance, in gene expression analysis, the PC is a linear combination of all genes, which is referred as mega genes, eigen genes, latent genes among others in published studies. The biological implication has not been fully understood. Meanwhile, it is more difficult to interpret such "mega compound" consisting of main and interaction effects in G × E studies as the two types of effects play distinct roles. Unsupervised analyses have been conducted for G × E interaction studies ([69−72]), but lack an extensive investigation. Indeed, as how the genetic variants are modified by environment factors to affect the risk of disease or variations of a trait is of the uttermost importance in G × E interaction studies, supervised techniques, such as variable selection, are more attractive than the unsupervised ones.

2.2 The form of environmental factors

In G × E interaction studies, the form of environmental exposures play a crucial role ([19]). Denote E and G as the environmental and genetic factors, respectively. From a statistically modelling perspective, G × E interactions in marginal analysis can be represented by the product between the two, as in $Y \sim \gamma E + \alpha G + \eta GE$, where regression coefficients $\gamma$ and $\alpha$ are the effects of the environment (E) and genetic (G) factors, respectively, and coefficient $\eta$ quantifies the effect of G × E interaction. Rearranging the expression yields: $Y \sim \gamma E + (\alpha + \eta E)G$, which clearly reveals that the contribution of a genetic variant to the variation in continuous disease trait is in the form of a linear function of the E factor, i.e.

$$\text{Outcome} \sim \text{E} + \text{G} \times (\text{linear function in E}), \quad (2)$$

where the main effect of the G factor is corresponding to the intercept of the linear function, therefore a separate term to model it is unnecessary. Linear interaction assumption has been widely adopted in a large number of G × E studies, especially statistical significance based ones such as GWAS. As multiple studies have shown that the interaction is not necessarily linear, Ma et al. ([73]) and Wu and Cui ([74]) are among the first to assess the nonlinear G × E interactions. In particular, Wu and Cui ([74]) has demonstrated that the model yields significant interactions in the form of SNP × (nonlinear function in BMI), which has not been captured by model (2) in two case-control studies of type 2 diabetes, the Health Professionals Follow up Study (HPFS) and the Nurses' Health Study (NHS), from the Gene, Environment Association Studies Consortium (GENEVA) ([75]). When the nonlinear interaction between G factor and low dimensional, say three, E factors is taken into account, it is natural to consider G × (nonlinear function in $U$), where the index function $U(= \beta_1 E_1 + \beta_2 E_2 + \beta_3 E_3)$ is an environmental mixture. An immediate extension from the above model is to consider both linear and nonlinear interaction in the same model. By far, we have only discussed marginal models where only one G factor is included. Recently, extensive efforts have been devoted to dissecting G × E interactions based on linear, nonlinear or both assumptions in the joint paradigm, especially using penalization methods. Figure 1 shows a taxonomy of G × E interaction studies reviewed throughout the article following the analysis framework, variable selection methods and conceptual models with diverse interaction assumptions. For simplicity of notation, clinical variables not involved in interactions, as shown in model (1), are dropped from all the conceptual models from now on.

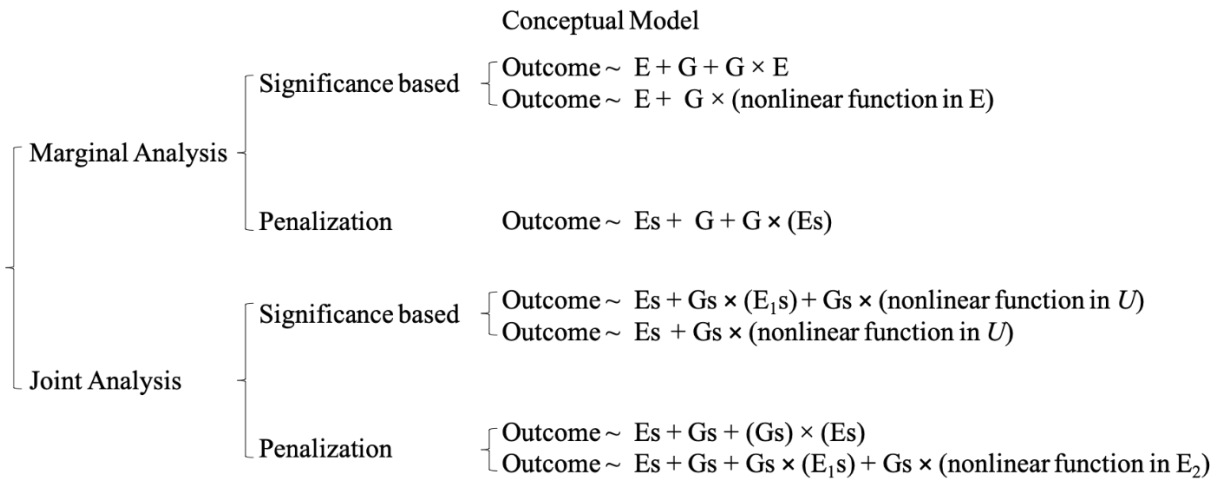

**Figure 1.** A taxonomy of conceptual models reviewed in marginal and joint analyses.

Following studies summarized throughout the paper and other recently published studies, we take a "loose" definition to treat clinical, prognostic and environmental risk factors as E factors. The G factor refers to any type of multi-omics data, including gene expression, SNP, methylation, and lipids, among many others. In general, one G/E factor is denoted as one variable in the model based framework. Special accommodations include considering both additive and dominant penetrance effect of the genetic variant, so two variables corresponding to the two effects are necessary to represent the genetic variant, as adopted in [73,76].

Joint analyses provide a strong contrasts between these two major types of genetic interaction studies. When high dimensional G factors, such as gene expressions, SNPs or other omics features, are taken into account, G × G studies investigate the interaction between two symmetric arrays of features where the number of features is much larger than sample size. The computational cost is daunting even merely for pairwise interactions, let alone the higher order epistasis. On the contrary, in majority of the G × E studies, the environmental factors are usually predetermined as important ones with evidence from published research. They are of low dimensionality and not subject to selection. Therefore, G × E interactions are more asymmetrical in that the interactions between a group of low dimensional E factors and a high dimensional G factors are of interest. Such an asymmetry, as a trademark of G × E studies, together with considerations of various forms of E factors, have generated extensive amount of novel, powerful and tailored penalization methods for G × E studies recently.

### 2.3 Other important considerations

We will also pay particular attention to the following aspects of the G × E interaction studies. First, penalized variable selection can be formulated in the form of "unpenalized loss function + penalty function". The penalty function determines how the selection on main and interaction effects is conducted. As the nature of interaction studies, such as linear or nonlinear interactions, characterizes the unpenalized loss function, the design of penalty function is under constraint. We will discuss how the tailored penalty function is developed based on the specific studies. Second, the demand for robustness in interaction studies arises since outliers and data contamination have been widely observed, mainly from response variables but also from predictors ([14,20]). Besides, model misspecification, such as mis-specifying the correct form of environmental factors (see discussions in Section 2.2), also needs to be remediated by robust interaction methods. Within the "unpenalized loss function + penalty function" framework, a robust loss function is necessary to incorporate robustness while selecting important effects, which distinguishes robust G × E interaction studies from the non-robust counterparts. Third, the hierarchical structure between main and interaction effects, including both strong and week hierarchy, is a unique characteristic of interaction studies. Strong hierarchy postulates that when G × E effect is identified, both the G and E effects should also be identified. The presence of only one of the two main effects upon identification of G × E interaction is referred as weak hierarchy. How to effectively accommodate the hierarchical structure has received much attention in recent development of variable selection methods for interaction analysis.

We organize the rest of the article based on marginal and joint analysis paradigms separately. Within each paradigm, we arrange the published studies according to increasing complexity levels of environmental factors. Discussions will be made if penalty functions, robustness and hierarchical structure are of particular interest for the G × E interaction studies.

2.4 Marginal Analysis

Since only a small fraction of G factors is taken into account each time, the dimensionality of variables with respect to the main and interaction effects is significantly lower than the sample size. Such an advantage guarantees that the marginal analysis is computationally convenient and stable. Therefore, such a paradigm is still popular nowadays, especially in large scale biomedical and bioinformatics studies.

2.4.1 Statistical significance based marginal studies

Methods in this category generally proceed as follows. For each G factor, set up a likelihood ratio test with the reduced model: Outcome ~ E + G, and full model: Outcome ~ E + G + G × E. Statistical test based on the two models yields p-value as the evidence of statistical significance for the interaction term. In large scale studies, such as the GWAS, the dimensionality of the G factors is ultra-high, usually on the magnitude of millions. Therefore, conducting the marginal tests for all the G factors will provide us millions of p-values. Based on the p-values and a properly chosen statistical significance threshold after correcting for multiple tests (Bonferroni, FDR, etc.), a Manhattan plot can then be generated to pinpoint the significant G × E interactions. With linear interaction assumption, such a marginal analysis has been widely conducted especially for binary traits in GWAS. A detailed review can be found from [10−12].

Next, we will specifically discuss the marginal analysis that is tightly connected to penalization methods. As the interaction might not necessarily be linear, statistical significance based variable selection methods have been examined by adopting the following conceptual model:

$$\text{Outcome} \sim E + G \times (\text{nonlinear function in E}), \quad (3)$$

where a separate term to model main genetic effect is not necessary as it has been absorbed into the nonlinear function as the intercept. Such a model advances from (2) by incorporating the contribution of the G factor to the response variable as a nonlinear function in the E factor. Ma et al. ([73]) and Wu and Cui ([74]) are among the first to conduct significance based interaction analysis with nonlinear G × E interaction assumptions, for continuous disease phenotype and case-control association studies respectively. With the null model: Outcome ~ E, p-values for testing whether there is a nonlinear interaction effects can be obtained. The two studies have performed sequential tests to determine whether the G factor (a) is associated with phenotype at all, (b) has linear interactions with the E factor and (c) has a nonlinear interaction with the E factor. It has been shown that for a dataset containing 1536 new born babies with 648 SNPs covering 189 genes, variations in birthweight (the phenotype) has been triggered by multiple nonlinear interactions between mothers' obesity condition (the E factor), measured as mother's body mass index (MBMI), and the infants' SNPs ([73]).

By far, our discussions are focused on the published marginal analyses where only one E factor is involved. When there are multiple E factors, it is possible that both linear and nonlinear interactions are present. Liu et al. ([77]) has proposed a joint test of linear and nonlinear effects in marginal G × E interactions. The joint test consists of a likelihood ratio test for linear interactions, and a generalized likelihood ratio test for nonlinear interactions. It has been reported that the joint test is better powered than alternatives that solely test only one of the two types of effects. A case study on the Thai fetal data from GENEVA shows how SNPs are mediated by mother's glucose level, represented by three E factors, to affect birthweight (the phenotype) in a linear and nonlinear fashion jointly.

2.4.2 Penalization based marginal studies

Consider the following conceptual marginal models

$$\text{Outcome} \sim \text{Es} + \text{G} + \text{G} \times (\text{Es}), \quad (4)$$

where Es denotes multiple E factors instead of a single one, and G × (Es) denotes the linear interaction between the G factor and all the E factors. Since the environmental factors are not subject to selection, the penalty function has only been imposed on the coefficients corresponding to the last two terms, the main and interaction effects involving G factor. Marginal penalization can be conducted one at a time for all the G factors, such as gene expressions and SNPs.

Robustness has been incorporated in marginal penalization to better accommodate data contamination and heavy-tailed errors in responses. Among these studies, Shi et al. ([78]) has proposed a robust semiparametric transformation model within the MRC (maximum rank correlation) framework for continuous, categorical and survival responses. The robustness is two folded. First, the rank based procedure is robust to contamination and heavy tailed distributions in the response variable. Second, the form of the link function in the semiparametric transformation model is unspecified, so robustness to model misspecification is guaranteed. Three independent lung cancer prognostic datasets with high dimensional gene expression measurements have been collected following [79]. Analysis of the heterogeneous data leads to interactions with important implications. Chai et al. ([80]) has also developed a robust variable selection method based on the same conceptual model. However, they chose an exponential squared loss function, which results in a completely different estimation procedures compared to Shi et al. ([78]). The lung squamous cell carcinoma (LUSC) data collected from TCGA has been analyzed to demonstrate the utility of the proposed model.

Imposing hierarchical structures through penalized identification of important main and interaction effects has recently been considered in many studies ([81,82]). With a G × E interaction term being identified, strong hierarchy is respected if both main genetic and environmental effects are also identified. A weak hierarchy indicates that at least one of the two main effects is present given the identification of interaction term. Since E factors are usually of low dimensionality and not subject to penalized selection, imposing strong hierarchy in G × E studies is more intriguing. Unlike the two-stage strategy which searches for important interactions and then force the corresponding main effects to stay in the model via refitting if not, such as in Chai et al. ([80]), developing penalization methods respecting hierarchy for G × E interactions is much more challenging.

Hierarchical penalized identification of main and interaction effects can be achieved through developing more complicated but equally efficient penalty functions. For instance, Zhang et al. ([83]) has proposed a marginal hierarchical penalization approach for G × E studies by adopting a bi-level sparse group penalty ([43,84]). It is equivalent to determining for one G factor, first, on the group level with all the G main effects and G × E interactions, whether the variant is associated with clinical outcome or not, and second, on the individual level (within the group), what are the specific effects that are associated with the response? The key to ensure strong hierarchy is to only penalize the coefficient of main G effect on the group level while all the interaction effects for the G factor are penalized on both levels. The main genetic effect thus has been shrunken with less amount compared to interaction effects. Therefore, once the interaction effects are identified, the main effect must also be present in the model. The strategy of imposing hierarchy has been shown to be effective first in joint models ([85,86]). Zhang et al. ([83]) is an important step towards borrowing strength from joint G × E analysis for the marginal ones to build a more coherent interaction analysis framework.

The formulation of model (4) has initially been investigated in significance based G × E interaction studies. Denote $G = (G_1, G_2, ..., G_p)$ as $p$ genes, and $E = (E_1, E_2, ..., E_q)$ as $q$ environmental factors. Then the most widely examined model based on the conceptual model (4) is $Y \sim \sum_{k=1}^{q} \gamma_k E_k + \alpha_j G_j + \sum_{k=1}^{q} \eta_{j,k} G_j E_k$ for $j = 1, ..., p$, where $Y$ is a continuous disease phenotype for simplicity, $\gamma_k$'s, $\alpha_j$ and $\eta_{j,k}$'s are unknown regression coefficients. For the $j$th G factor, the coefficient vector $(\alpha_j, \eta_{j,1}, \eta_{j,2}, ..., \eta_{j,q})$ represents main genetic and $q$ G × E interactions. Fit the $p$ regression model using standard procedures and software to obtain p-values of the estimates of $\alpha_j$ and $\eta_{j,k}$. Then significant main and interaction effects can be identified after adjusting for multiple comparison with Bonferroni or false discovery rate (FDR). Recently, Xu et al. ([87]) has proposed a robust censored quantile partial correlation (CQPCorr) technique based on (4). The p-values yielded from CQPCorr through permutation can be adopted for significance based identification of main and interaction effects.

**Remarks on the choices of penalty functions under model (4)**: "Rank based loss + MCP" and "Exponential squared loss + LASSO" have been adopted for Shi et al. ([78]) and Chai et al. ([80]), respectively. Smoothed approximations to the robust loss functions have been conducted to guarantee the compatibility in terms of computation between the loss and penalty functions. Here, the choice of penalty function is more flexible as only the baseline level penalty, such as LASSO and MCP, is needed, and more attention is paid to loss functions as smoothed approximation is necessary. The strong hierarchy is respected in Chai et al. ([80]) though refitting the main G factor if it's not selected after penalization. Zhang et al. ([83]) has adopted sparse group MCP, a bi-level penalty ([43]), to impose a natural strong hierarchy without refitting. In general, the structured sparsity can be more effectively accommodated under more tailored penalty functions.

## 2.5 Joint Analysis

### 2.5.1 Statistical significance based joint studies

Simultaneous marginal tests on both linear and nonlinear interaction effects have been conducted in Liu et al. ([77]). It is more challenging to develop joint tests for the following conceptual model:

$$\text{Outcome} \sim \text{Es} + \text{Gs} \times (\text{E}_1\text{s}) + \text{Gs} \times (\text{nonlinear function in } U), \quad (5)$$

where the index $U$ is a mixture of environmental factors $\text{E}_2\text{s}$, i.e. $U = \beta^T(E_2\text{s})$, with the loading parameter vector $\beta$. E = ($\text{E}_1\text{s}$, $\text{E}_2\text{s}$) consists of two sets of environment factors interacting with multiple genetic factors (Gs) linearly and nonlinearly, respectively. A separate term to model main effect of the G factor is not needed since it has been incorporated as the intercept of the nonlinear function. Compared to the model (3), the complexity of G × E interactions has increased as the index of multiple environmental factors, $U$, instead of a single E variable, is involved in the nonlinear interaction with G factors.

A simple version of model (5) has been first developed. Ma and Xu ([88]) has conducted an exploratory analysis on data from Framingham Heart Study to motivate the partially linear single-index coefficient model. Strong nonlinear interaction pattern can be observed using BMI as response and three time-related covariates (sleeping hours per day, hours of light activity per day and hours of moderate activity per day) as E factors to form index $U$. The study proposed by Ma and Xu ([88]) is a special case of the model (5) as follows:

$$\text{Outcome} \sim \text{Es} + \text{Gs} \times (\text{nonlinear function in } U),$$

which drops the linear interaction terms in model (5) and captures the main genetic effects in the nonlinear function so an additional term for G factor is not necessary. The clinical covariates besides the E factors are also included, and omitted here for simplicity of notation. Score tests have been proposed to test the significance of the linear effects and nonlinear G × E interactions. Note that "linear" in Ma and Xu ([88]) only refers to the main effects of environmental and the additional clinical factors, not the linear G × E interactions. Real data analysis reveals linear effects and nonlinear G × E interactions on body mass index based on four SNPs, three aforementioned E factors and five clinical covariates from the Framingham Heart Study.

The model (5) has been fully explored in Liu et al. ([89]), where both linear and nonlinear G × E interactions are tested in a joint fashion. The study has also considered different index loading vectors instead of assuming the same one ($\beta$) for different G factors as adopted in Ma and Xu ([88]). A case study of six SNPs, one discrete E factor (infants' gender) in $\text{E}_1$, three continuous E factors (related to mother's glucose level and blood pressure) in $\text{E}_2\text{s}$ based on the Thai population from GENEVA has shown how genetic effects interacts with gender ($\text{E}_1$) linearly and environmental mixtures ($\text{E}_2\text{s}$) nonlinearly to influence birthweight.

In this section, we survey statistical significance based joint studies that are strongly connected to penalization methods, as shown in Section 2.4.2. Other joint testing based G × E studies, including [90, 91], have also been conducted but are less relevant to penalization. Thus they are not discussed here.

Under model (5), the E factors are in the form of environmental mixtures. The index function is defined as $U = \beta^T(E_2s)$. To illustrate, with three environmental factors (for example, as used in Ma and Xu [88]) in $E_2$s, i.e., $E_{21}, E_{22}$ and $E_{23}$, it can be expressed as $U = \beta_1 E_{21} + \beta_2 E_{22} + \beta_3 E_{23}$, where $\beta = (\beta_2, \beta_2, \beta_3)^T$ is the loading vector. Although the complexity of E factors dramatically increases in the form of mixture of environmental factors, the utility of the model has been convincingly demonstrated in [88, 89] among others. Liu et al. ([77]) has relaxed the single index assumption in Ma and Xu ([88]) and allowed different loading vectors for different G factors as a multi-index model. In addition to G × E studies, interactions involving the environmental mixtures have recently been examined using sparse Bayesian methods ([92,93,94]). In these studies, environmental mixtures refer to the complex (such as nonlinear, and/or high order) interactions among low dimensional E factors, so the dimensionality is still high—just imagine the total dimensionality of three-way interactions from 10 E factors. These studies essentially share the spirit of G × G interactions and are different from our discussions in G × E studies.

**Remarks on a unified framework of varying index coefficient models**: Ma and Song ([95]) has provided a detailed introduction on the interconnections among a broad spectrum of existing models within the framework of varying index coefficient models, including single index coefficient model, nonparametric additive model, partially linear single index model among many others. Penalized variable selection has already been developed for single and multi-index models. However, a common limitation is that these studies only examine variable selection in very low dimensions where number of predictors is much less than the sample size, partially due to the expensive computational cost. From perspective of statistical hypothesis testing, we can also observe this limitation from [88,89], which conduct marginal screening to reduce the number of SNPs in the final analyses to 4 and 6, respectively.

2.5.2 Penalization based joint studies

With multiple G factors and the linear interaction assumption, the following conceptual model serves as an umbrella framework for a large number of G × E interaction studies:

$$\text{Outcome} \sim \text{ Es} + \text{ Gs} + (\text{Gs}) \times (\text{Es}), \quad (6)$$

where Es and Gs denote multiple environmental and genetic variables. The E factors are low dimensional and pre-chosen from following published studies. It is an extension from the marginal model (4). As discussed, such a formulation naturally leads to a sparse group (or bi-level) selection problem. With respect to each G factor, the group consists of main G and G × E interactions in the form of G × (1, Es). The sparse group variable selection determines, on the group level, if the genetic variant is associated at phenotype. If not, the regression coefficients corresponding to the group of main and interaction effects are all zero. If yes, then on the individual level within the group, what are the subset of important effects that are associated with the phenotype? A common strategy of penalized identification is to impose both the group level penalty and individual level penalty using the baseline penalty functions, such as LASSO, adaptive LASSO, SCAD, MCP and others, to form sparse group LASSO/Adaptive LASSO/SCAD/MCP among others.

This model has initially been developed in a study of sparse group MCP for hierarchical bi-level selection ([85]). To respect strong hierarchy, the main effect is only subject to penalization on the group level, and all the interaction effects are penalized on both group and individual levels. So the identification of interaction effects ensures the inclusion of G main effect since the coefficient of main effect receives less amount of shrinkage. This strategy has also been adopted in Wu et al. ([86]) and the previously discussed Zhang et al. ([83]). Liu et al. ([85]) has extended the model from a continuous response to the right censored survival response under the accelerated failure time (AFT) model, resulting in a penalized least square loss function reweighted based on the Kaplan-Meier estimator, and a similar estimation procedure as for the continuous responses. The non-Hodgkin lymphoma (NHL) data with SNP measurements collected from [96,97] has been analyzed using the proposed method with four E factors age, education level, tumor stage and initial treatment. Promising identification and prediction results have been observed through the case study.

A further improvement over Liu et al. ([85]) is conducted through incorporating robustness in the hierarchical identification of G × E structures to tackle skewed cancer prognostic outcomes ([86]). Wu et al. ([86]) has adopted the least absolute deviation (LAD) loss in the study mainly due to computational simplicity. As a special case of the check loss in quantile regression, LAD loss has been extensively examined in low dimensional settings and high dimensional main effect models. Robust loss function usually demands much more complicated computational algorithms, and such complexity scales up when structured variable selection is of interest. It has been further demonstrated that Kaplan–Meier weights can be utilized to accommodate hierarchical G × E interactions efficiently and robustly in cancer prognostic studies. A case study of non–small-cell lung cancer data collected from four independent studies ([79]) yields superior predictive power over alternatives (without imposing hierarchy and/or robustness) to partially justify the advantage of the proposed robust hierarchical penalization method.

The interconnections among genomics features widely exist, and have motivated the development of a large amount of penalization methods, including elastic net, fused LASSO and network constrained variable selection. In G × E studies, accounting for the relatedness of genomics variants has been first considered in Wu et al. ([98]) under model (6). In addition to the sparsity induced penalty functions on the main and interaction effects, Wu et al. ([98]) has adopted a spline type of penalty to promote smoothness among densely measured G factors, such as SNPs, and a network-constrained penalty to encourage correlations among G factors. In addition, regression coefficients have been reparametrized to force hierarchical structure, which leads to a different mechanism to incorporate strong hierarchy from [85,86]. The effectiveness of the G × E method has been shown on both a type 2 diabetes study with high dimensional SNP data from GENEVA and TCGA melanoma data with high dimensional gene expression measurements.

Wang et al. ([99]) has conducted a very interesting G × E interaction study by utilizing prior information on associations between G factors and the disease phenotype from published studies. In particular, the G factors in (6) have been decomposed into two components, according to whether they have been reported to be associated with the disease. A tuning parameter is chosen data dependently as a weight to balance the two components in penalized objective function. Such a formulation has motivated a quasi-likelihood based estimation procedure. Detailed steps of how to use the online tool *PubMatrix* to mine prior information from published studies has been provided for the analyses of TCGA skin cutaneous melanoma (SKCM) and glioblastoma multiforme (GBM) data.

The types of phenotype from studies we surveyed so far are mostly continuous and survival. Recently, interaction studies have been extended to longitudinal traits. Zhou et al. ([100,101]) have developed a generalized estimation equation (GEE) based penalization method to identify lipid-environment interactions in a cancer prevention study with repeated measurements of weight and lipids from CD-1 mice. The group of E factors consists of three dummy variables built on a treatment related to dietary restriction and exercises with four levels. Only group level penalty is imposed on Lipid × Environment interactions to enforce group-in or group-out selection. Individual level penalty is only imposed for selecting main lipid effects. So different from the bi-level selection, main G and G × E effects are merely subject to selection on group and individual levels respectively instead of simultaneously. The superior performance of proposed methods over alternatives in terms of identification, prediction, scalability and stability has been fully demonstrated. Furthermore, Zhou et al. ([102]) have further relaxed the restriction of group-in/group-out selection on E factors to examine a more general sparse group variable selection in longitudinal studies using quadratic inference functions.

While all the discussed interaction studies based on model (6) are proposed within frequentist framework, Ren et al. ([103]) has investigated robust Bayesian variable selection under this umbrella model. A set of innovative robust Bayesian methods accommodating the sparse group, group and individual level structure have been developed with and without using spike-and-slab priors. Proposed and benchmark methods have been applied on SNP data from a type 2 diabetes study and TCGA melanoma study. This work significantly advances from existing Bayesian variable selection methods in that both robustness and structural sparsity (bi-level selection) have been considered.

The list of interaction studies within the umbrella framework outlined by (6) keeps growing recently. For example, Wu et al. ([104]) has shed new insight into the model by accommodating missingness in environmental measurements. Also, Du et al. ([105]) has conducted integrating multi-omics data for gene-environment interactions. Xu et al. ([106]) has investigated penalized trimmed regression in G × E studies. Xu et al. ([107]) has examined imaging-environment interactions based on the conceptual model in both marginal and joint studies.

**Remarks on the choices of penalty functions under model (6)**: The interplay between loss and penalty functions within the "unpenalized loss function + penalty functions" framework has been further revealed by comparing Wu et al. ([86]) with the marginal analyses ([78,80]). To accommodate a large number of main and interaction effects in one joint model, the LAD loss of a simple L1 form has been chosen due to computational considerations. To make the sparse group adaptive LASSO penalty consistent with the L1 form of the loss function, a smoothed approximation has been conducted on the group adaptive LASSO penalty which is of the square root form, so eventually we only need to deal with LAD based computation. Compared with [78,80], Wu et al. ([86]) has a simpler loss function but more complicated penalty structure (so adjustment has been made on the penalty), whereas a more complicated loss but simpler baseline penalty (MCP and LASSO) have been adopted in [78,80]. Such a difference is caused by the fact that baseline penalty cannot directly lead to strong hierarchy. When a relatively complicated sparse group penalty function is adopted ([86]), the unpenalized loss function is supposed to be convenient to facilitate efficient computation in joint analysis. Besides, the lipid study of Zhou et al. ([100]) again manifests the special role of E factors in G × E studies. By construction, the E factor shares the spirit of group LASSO, where only group level penalty, instead of both group and individual level penalty as in [85,86,103], is required for interaction effects.

The model (6) assumes linear G × E interactions. Next, we turn our attention to a conceptual model incorporating both linear and nonlinear interactions:

$$\text{Outcome} \sim \text{Es} + \text{Gs} \times (E_1s) + \text{Gs} \times (\text{nonlinear function in } E_2s), \quad (7)$$

where $E_1s$ and $E_2s$ are environment factors that interact with G factors (Gs) linearly and nonlinearly, respectively. Combined, E = ($E_1s$, $E_2s$) denotes the all the preselected E factors. Again, note that the main genetic effect of the G factor is captured by the nonlinear function as intercept, so a separate term to model the main effect is not necessary.

Model (7) has deep connections to previously discussed ones. For example, in the marginal framework, with only one G and one E factor, model (7) reduces to model (3) (or model (2)) through dropping the linear (or nonlinear) interaction term. Furthermore, with one G and multiple E factors, models (7) and (4) are equivalent when the nonlinear interaction in model (7) vanishes. The connection between model (6) and (7) can be established similarly.

Migrations of marginal nonlinear G × E methods ([73,74]) to joint paradigms have been primarily motivated by gene set and pathway based association analysis ([9,108,109, 110]). A simplification of conceptual model (7) of the following form has been first proposed along the line:

$$\text{Outcome} \sim \text{E} + \text{Gs} \times (\text{nonlinear function in E}), \quad (8)$$

where the nonlinear interaction between multiple G factors and one environmental factor is considered. In G × E interaction studies, such a model has been first developed in [111,112] with structural sparsity. As previously discussed, the nonlinear function in E factor incorporates both the genetic main effect and G × E interactions, so the nonlinear function reduces to: (a) zero if there's no genetic association at all; (b) a nonzero constant if only genetic main effect exists. The method proposed in [111,112] is capable of accommodating potential structural misspecification issues. SNPs interacting with the E factor (Mother's BMI or MBMI) using the dataset containing 1536 infants with 660 SNPs from 189 genes have been identified under the phenotype birthweight. Such a model has also been examined in [49,113], among many other studies, however, without imposing structural identification.

The conceptual model has been adopted and significantly enriched in integrative G × E analysis ([114]) in the following aspects. First, as the nonlinear function of $E_2s$ incorporates both G main effect and G × E interactions, to avoid model misspecification, Wu et al. [114] conduct a structural identification by separating the main effect from G × E interactions, so the two types of effects are handled separately. Missing such a step will yield inaccurate identification and prediction results especially when only one of the two effects exists. Second, the multivariate response and two model assumptions have been considered ([114]). With homogeneity model assumption, the same set of main and interaction effects are associated with all the traits, which is consistent with the phenomena of pleiotropy in which common genetic basis is shared among correlated traits. On the contrary, heterogeneity model assumption does not have this restriction, and different sets of effects can be associated with different traits. Such complicated data and model settings leads to different penalization mechanisms. In the case study, homogeneity model assumption yields better results under two positively correlated traits, BMI and weight, with age (for nonlinear interaction) and family history of diabetes among first degree relatives (for linear interaction) as the environmental factors and SNPs as the genetic factor, from the Health Professionals Follow-up Study (HPFS) — a GWAS launched by GENEVA.

Robustness can be incorporated into G × E studies assuming not only linear interactions such as in model (6), but also both linear and nonlinear interactions. Wu et al. ([115]) has proposed a robust semiparametric model under (7). Here, the robustness is two folded. First, a rank based loss function is adopted to take care of outliers and data contamination in phenotype. Such a loss function can be approximated by a reweighted least square loss, which dramatically reduces computational cost given the non-differentiable rank-based loss function and a large number of main and interaction effects. Second, structural identification by separating main and interaction effects from the nonlinear interactions has also been imposed, which is robust to model misspecification. A prognostic study of gene expression data collected from four independent studies on non–small-cell lung cancer ([79]) identifies important interaction effects, both linearly (with smoking status) and nonlinearly (with age). Multiple main genetic effects have also been detected. Without the structural sparsity induced penalty function, these main G effects will be identified as nonlinear interactions. The real data analysis again manifests the motivation of developing robust G × E methods.

Penalization and Bayesian shrinkage variable selection are two sides of the same coin. As penalization methods have been quite successfully proposed for interaction studies in the past a few years, a promising trend nowadays is to conduct Bayesian penalized variable selection to gain fresh insight into the study. Ren et al. ([50]) has developed a semiparametric Bayesian variable selection method based on model (7). Structural identification by separating the genetic main effect and G × E interactions leads to the selection on both group (nonlinear G × E interactions) and individual levels (G main effect and linear G × E interactions). The simultaneous Laplacian shrinkage with spike-and-slab priors on both group and individual level significantly advances from existing Bayesian methods for G × E interactions. More importantly, it has also been demonstrated that the proposed method is scalable under even extreme "large $p$, small $n$" scenarios. The computational efficiency is partially due to a C++ implementation of the Gibbs sampler, which is available from R package *spinBayes* ([116]). The proposed Bayesian method has resulted in findings with important biological implications using age and the binary indicator of whether an individual has a history of hypertension (hbp) or not as the E factor for linear and nonlinear interactions respectively, SNP as the G factor, and weight as the phenotype.

In Section 2.1, we have discussed the connection between penalization and traditional machine learning methods, including Boosting. Such a connection is the driving force of development of new boosting methods for detecting G × E interactions. As increasing amount of attention has been paid to robustness and hierarchical structure of interaction studies, Wu and Ma ([61]) has proposed a robust semiparametric sparse Boosting approach for simultaneous identification of both linear and nonlinear interactions. They have adopted the Huber loss function for robustness, and a multiple imputation approach to accommodate missing values in the E factor. In the case study of TCGA skin cutaneous melanoma (SKCM) data with gene expression as the G factor, three different groups of nonlinear interactions, between G and age, weight and height respectively, are taken into account, while all the rest of the studies discussed here under model (8) consider the nonlinear interaction in only one E factor. Also note that model (8) can be treated as a simplified specification of varying index coefficient models with only one E factor in the environmental mixtures $U$, as we have mentioned at Section 2.5.1 ([95]).

Without the main environment effect, model (8) reduces to the classical varying coefficient models ([117,118]) where the coefficient of covariates are allowed to fluctuate in some other variables via smooth functions in order to assess nonlinear interactions. Variable selection in varying coefficient models, as well as their variants, have been extensively investigated. However, limited efforts have been devoted to connecting the model to G × E studies before [73, 74, 76, 111, 112]. One of the consequences is, the case studies from majority of these literature are restricted to data with small size such that the number of variables subject to selection is usually on the magnitude of 10 only, even though the study has been proposed for "high dimensional" variable selection. For example, The HIV infection data from the multicenter AIDS cohort study ([119]) conducted over 30 years ago have still been used repeatedly to assess the nonlinear time-varying effects of low dimensional covariates, including smoking status, pre-HIV infection CD4 cell percentage and age at HIV infection, on the response CD4 percentage. The G × E interaction studies have revived varying coefficient (and relevant) models from a real high dimensionality perspective.

**Remarks on the choices of penalty functions under model (7)**: As nonlinear interactions are usually modeled through smooth functions, nonparametric estimation procedures, such as those based on splines or kernels ([118]), are possible choices. The spline based methods are especially appealing due to computational efficiency with high dimensional data. Through basis expansion using splines (B spline or polynomial spline, among many others), identification of the nonlinear interaction is equivalent to the selection of a group of basis functions ([120]), which eventually leads to a group level selection problem ([42]). Therefore group level penalty, such as group LASSO/SCAD/MCP, are generally demanded. Variations include structural identification by separating the main effect from the nonlinear interactions, which motivates a combination of individual level penalty (on the genetic main effect) and group level penalty (on the nonlinear interactions).

2.6 Computational aspects in G × E studies

Our survey shows that coordinate descent (CD) is the most popular computational framework for penalization based G × E interaction studies, especially in the joint paradigm. The penalized loss function, which in the form of "unpenalized loss function + penalty function", is optimized with respect to one predictor (or predictor group) at a time across all the coordinates till convergence ([121−123]). The computation within CD is particularly fast when first order methods, which are generally based on gradient, sub gradient and proximal gradient, have been developed to tackle a wide variety of optimization problems under both non-robust and robust loss functions with penalty functions designed for different studies ([124,125]). Our remarks on the choices of penalty functions in G × E studies have revealed that penalty functions for the individual level, group level, and a combination of the two levels (or bi-level) are the most popular ones in G × E studies, and all of them can be efficiently accommodated within the CD framework. For example, in semiparametric interaction studies ([114,115]), the shrinkage has been imposed on both the individual level (main genetic effect and linear G × E interactions) and group level (nonlinear G × E interactions), then the optimization within CD can be conducted with respect to a mixture of individual and group coordinates one at a time until convergence. Other efficient algorithms, including those such as the alternating direction method of multipliers (ADMM), least angle regression (LARS), iterative shrinkage thresholding algorithm (ISTA) ([126,127,128]) among many others, have also been developed for penalization methods, however, they are less popular in G × E studies.

The formulation of "unpenalized loss function + penalty function" sheds fresh insight into the G × E interaction studies from a variable selection point of view. The nature of the studies shapes the loss function, which may in turn pose restrictions or demand modifications on the penalty function (please refer to our remarks on the choices of penalty functions). Furthermore, the choices of penalty functions are also dependent on the structure of omics measurements. For example, fused LASSO has been developed to account for spatial correlations across neighboring genomic regions in CNVs and other omics features ([129]). The network-based penalties have been proposed to incorporate correlations using a network, which are more general and not restricted to neighboring genomic positions ([39, 40, 130−132]). Wu et al. ([98]) is among the first to accommodate strong interconnections in G factors through networks in G × E studies, while such a strategy is more frequently observed in epistasis studies ([27,133−135]).

Bayesian variable selection methods have also been developed to identify important G × E interaction effects. Within the Bayesian framework, the appropriate specification of hierarchical model with suitable prior distributions can lead to the Gibbs sampler, generating posterior samples efficiently from full conditional distributions. For example, the Laplacian shrinkage via spike-and-slab priors have been imposed on both the group level for nonlinear interactions and the individual level for main G and linear G × E effects in Ren et al. ([50]), which amounts to a Gibbs sampler to guarantee fast convergence to stationary distributions and effective identification of important effects. A Metropolis Hastings step is necessary when sampling from full conditional is not feasible ([49]).

Our summary in Section 2.1 shows that besides penalization and Bayesian methods, other popular machine learning methods have also been proposed. For example, the strategy of using a stage-wise approach to progressively optimize the objective function has been adopted in Boosting based G × E studies ([60, 61]).

**Table 2.** Published G × E interaction studies using variable selection methods (a partial list).

| Method | Formulation | Data | Software |
| --- | --- | --- | --- |
| Shi et al. ([78]) | Marginal/Robust<br><br>Rank + MCP | Lung cancer data from Xie et al. ([79]). Y: patients' survival; G: gene expressions (sample size $n$=336 and dimension $p$=2500); E: age, gender, smoke, chemo, stage. | |
| Chai et al. ([80]) | Robust/Marginal/hierarchy<br><br>Exponential + L1 penalty | TCGA LUSC Data. Y: patients' survival; G: gene expression ($n$=404 and $p$=18,969); E: age, gender, smoking level, smoking status. | |
| Zhang et al. ([83]) | Marginal/Hierarchy<br><br>LS/survival + sparse group MCP | (1) GENEVA Type 2 diabetes. Y: BMI; G: SNPs ($n$=2,558 and $p$=10,000); E: age, famdb, act, trans, ceraf, heme.<br><br>(2) TCGA skin cutaneous melanoma. Y: patients' survival; G: gene expressions ($n$=298 and $p$=10,000); E: age, PN, gender, breslow's depth, clark level. | |
| Liu et al. ([85]) | Joint/Robust/Hierarchy<br><br>LS/survival + sparse group MCP | NHL Prognosis study. Y: patients' survival; G: tag SNPs from candidate genes ($n$=346 and $p$=1,633); E: age, education level, tumor stage and initial treatment. | https://github.com/<br><br>shuanggema/sparse_gmcp |
| Wu et al. ([86]) | Joint/Robust/Hierarchy<br><br>L1 + sparse group adaptive LASSO | Lung cancer data from Xie et al. ([79]). Y: patients' survival; G: gene expressions ($n$=351 and $p$=500). E: Age, Gender, Smoke, Chemo, Stage. | https://github.com/cenwu<br><br>/RobustHierGXE |
| Wang et al. ([99]) | Joint<br><br>Weighted quasi-likelihood + sparse group MCP | (1) TCGA skin cutaneous melanoma. Y: patients' survival; G: GEs ($n$=294 and $p$=1,350); E: age, stage, gender, clark level.<br><br>(2) TCGA GBM data. Y: patients' survival; G: GEs ($n$=300 and $p$=1,314); E: age, gender, KPS, race. | |
| Zhou et al ([100]) | Joint/robust<br><br>GEE+MCP+ Sparse group MCP | Lipidomics study from King et al. ([136]). Y: weight; G: lipids ($n$=351 and $p$=31); E: a group of three dummy variables formed based on the treatment (control, AE, PE and DCR) related to exercise and/or dietary restriction. | R package *interep* |
| Li et al. ([49]) | Joint<br><br>LS+ Bayesian group LASSO | Framingham Heart Study Y: BMI; G: SNPs ($n$=493 males and 372 females, $p$=33,239); E: age (nonlinear). | |
| Wu et al. ([114]) | Joint<br><br>Multivariate+ LASSO + group LASSO | GENEVA Type 2 diabetes. Y: weight and BMI; G: SNPs ($n$=2,568 and $p$=388) E: age (nonlinear) and FADMB (linear). | Github* |
| Wu et al. ([115]) | Joint/robust<br><br>Rank+ LASSO + Group LASSO | Lung cancer data from Xie et al. ([79]). Y: patients' survival; G: gene expressions ($n$=351 and $p$=200); E: Age(nonlinear) and smoking status. | Github* |
| Ren et al. ([50]) | Joint,<br><br>LS+ Bayesian LASSO + Bayesian group LASSO | GENEVA Type 2 diabetes. Y: weight and BMI; G: SNPs ($n$=1,716 and $p$=269); E: age (nonlinear) and HBP. | R package *spinBayes* |

| Wu and Ma ([61]) | Joint/Robust/hierarchy  Huber loss + sparse boosting | TCGA (stomach (gastric) adenocarcinoma) STAD data. Y: overall survival. G: gene expressions (*n*=381 and *p*=2,000). E: age (nonlinear), PM, PN, PT, gender, ICDO3 histology, ICD O3 site, and history of other malignancy. | Github* |
|---|---|---|---|
| | | TCGA SKCM data. Y: log transformed Breslow's depth; G: gene expressions (*n*=340 and *p*=2,000); E factor: age(nonlinear) weight (nonlinear), height (nonlinear), clark level, PN, PT, and sample type. | |

\* The corresponding authors' Github webpage
\* The G × E interactions are linear unless specified as nonlinear.

3. Notes

   1. We have provided a concise overview of penalization and relevant variable selection methods for gene-environment interaction studies under a variety of conceptual models. Such an effort has not been made in existing reviews on genetic interaction studies, including both G × G and G × E interaction. Our survey has investigated existing gene-environment interaction analyses from a fresh perspective based on variable selection. A price paid is that a large variety of important methods developed for interaction analyses has been inevitably left out from our study.
   2. It has been recognized that environmental factors in the rigorous sense are almost not available in TCGA. However, multiple G × E interaction studies, as shown from Table 2 among many others, have convincingly demonstrated the utility of analyzing TCGA data in interaction studies and yielding promising findings via taking a "loose" definition on E factors. Overall, variable selection stands out as a promising tool for interaction studies.

Acknowledgement

We thank the editor and reviewers for their invitation, careful review and insightful comments, leading to a significant improvement of this article. This study has been partly supported by the National Institutes of Health (CA191383, CA204120), the VA Cooperative Studies Program of the Department of VA, Office of Research and Development, an innovative research award from KSU Johnson Cancer Research Center and a KSU Faculty Enhancement Award.